\documentclass[reprint]{nst}

\usepackage{subfigure,dcolumn}
\usepackage{epstopdf}
\usepackage{mhchem}
\usepackage{upgreek}
\usepackage[normalem]{ulem}

\begin{document}

\title{Realistic Detector Geometry Modeling and Its Impact on Event Reconstruction in JUNO}

\author{Zhaoxiang Wu}
\affiliation{Institute of High Energy Physics, Chinese Academy of Sciences, Beijing 100049, China}
\affiliation{School of Physical Sciences, University of Chinese Academy of Sciences, Beijing 100049, China}

\author{Miao He}
\email[Corresponding author, ]{hem@ihep.ac.cn}
\affiliation{Institute of High Energy Physics, Chinese Academy of Sciences, Beijing 100049, China}

\author{Wuming Luo}
\email[Corresponding author, ]{luowm@ihep.ac.cn}
\affiliation{Institute of High Energy Physics, Chinese Academy of Sciences, Beijing 100049, China}

\author{Ziyan Deng}
\affiliation{Institute of High Energy Physics, Chinese Academy of Sciences, Beijing 100049, China} 

\author{Wei He}
\affiliation{Institute of High Energy Physics, Chinese Academy of Sciences, Beijing 100049, China}

\author{Yuekun Heng}
\affiliation{Institute of High Energy Physics, Chinese Academy of Sciences, Beijing 100049, China}

\author{Xiaoping Jing}
\affiliation{Institute of High Energy Physics, Chinese Academy of Sciences, Beijing 100049, China}

\author{Bo Li}
\affiliation{Institute of High Energy Physics, Chinese Academy of Sciences, Beijing 100049, China}

\author{Xiaoyan Ma}
\affiliation{Institute of High Energy Physics, Chinese Academy of Sciences, Beijing 100049, China}

\author{Xiaohui Qian}
\affiliation{Institute of High Energy Physics, Chinese Academy of Sciences, Beijing 100049, China}

\author{Zhonghua Qin}
\affiliation{Institute of High Energy Physics, Chinese Academy of Sciences, Beijing 100049, China}

\author{Yifang Wang}
\affiliation{Institute of High Energy Physics, Chinese Academy of Sciences, Beijing 100049, China}

\author{Peidong Yu}
\affiliation{Institute of High Energy Physics, Chinese Academy of Sciences, Beijing 100049, China}
\affiliation{School of Physical Sciences, University of Chinese Academy of Sciences, Beijing 100049, China}

\begin{abstract}
JUNO is designed to determine the neutrino mass ordering with an energy resolution of 3\% at 1 MeV. In the real detector, however, deformations of the central stainless-steel structure during installation lead to deviations of the photomultiplier tube (PMT) positions from their design values. Based on the limited survey data of the PMTs and the stainless-steel truss, we perform a correlation analysis of the measured points and propose a method to predict the positions of all PMTs. Using the resulting realistic geometry, we demonstrate that the detector deformation has a negligible effect on the energy reconstruction. In contrast, inaccuracies in the assumed geometry can introduce vertex 
biases of up to 40 mm. Incorporating the realistic geometry into the calibration-based PMT response model removes this bias and preserves the stability of the reconstruction algorithms.
\end{abstract}


\maketitle

\section{Introduction}
Neutrino oscillations provide compelling evidence of a non-zero neutrino mass, marking the first experimental observation~\cite{Super-Kamiokande:1998kpq, SNO:2001kpb, SNO:2002tuh} beyond the Standard Model. Despite this breakthrough, the absolute magnitude of neutrino masses and their mass ordering remain unresolved. The Jiangmen Underground Neutrino Observatory (JUNO)~\cite{JUNO:PPNP,JUNO:CDR} is a state-of-the-art 20-kton liquid scintillator detector located 700 meters underground, equipped with 17,612 20-inch PMTs~\cite{JUNO:2022hlz} and 25,600 3-inch PMTs~\cite{sPMT, JUNO:2025dfn}.

JUNO aims to determine the neutrino mass ordering by precisely measuring the energy spectrum of reactor antineutrinos~\cite{Zhan:2008id, Zhan:2009rs, Li:2013zyd}. Achieving this ambitious goal requires an unprecedented energy resolution of 3\% at 1 MeV~\cite{JUNO:yellowbook}. To meet this stringent requirement, JUNO employs PMTs with world-leading quantum efficiency~\cite{JUNO:2022hlz} to maximize photon detection, and extensive research has been conducted to optimize the liquid scintillator (LS) recipe for enhanced light yield~\cite{LSpaper}. Additionally, advanced algorithms have been developed to simultaneously reconstruct the event vertex and energy using charge and time information from PMTs~\cite{Huang:2021baf, Wu_2019, Liu_2018, Li_2021, Huang:2022zum}. Recent studies indicate that these algorithms perform well in reconstructing simulated positron events, achieving an energy resolution of 2.95\% at 1 MeV~\cite{JUNO:2024fdc}. Furthermore, simulations suggest that JUNO will achieve a median sensitivity of 3-4 $\sigma$ for determining the neutrino mass ordering after 6.5 years of data collection~\cite{JUNO:2024jaw}.

In previous Monte Carlo (MC) studies, the detector geometry is assumed to precisely match the design specifications, and this idealized geometry is consistently used in both detector simulation and event reconstruction. However, this assumption does not hold in the real experiment due to unavoidable deformations of the detector during installation. Specifically, the PMTs are designed to be distributed on a spherical surface, and any deviations from their intended positions can introduce biases in the charge and time predictions used in the reconstruction algorithm. These biases might degrade the resolution of event energy and vertex. The coordinates of all PMTs are crucial inputs to the event reconstruction of reactor neutrinos in the MeV region~\cite{Qian:2021vnh,Li:2022tvg,Jiang:2024wph,Gavrikov:2022kok}, as well as to the reconstruction and classification of cosmic muons and atmospheric neutrinos in the GeV region~\cite{Yang:2022din,Yang:2023rbg,Liu:2025fry}. Moreover, they are an indispensable component of the MC simulation for the JUNO central detector (CD)~\cite{Lin:2016vua,Zhu:2018mzu,Zhang:2020jkg}. Thus, one must obtain the most realistic and accurate PMT positions. 

To address this issue, several surveys of the detector have been conducted during the installation phase to accurately map the locations of the 20-inch PMTs. However, in order to minimize the interference with the detector installation, only a small number of positions were measured during the surveys. Given the enormous size of the JUNO CD, large number of PMTs, and the potential irregular deformation, it is quite challenging to obtain a realistic yet precise geometry based on these limited survey data. In this paper, we propose a method to model the PMTs' positions using a limited number of measured points obtained in the early survey stage, and consequently check the impact of the realistic geometry on the performance of the event reconstruction. An analysis of the survey data is described in Sec.~\ref{sec:survey}. The modeling of PMT positions is introduced in Sec.~\ref{sec:realization}. The impact of the detector geometry on the event vertex and energy reconstruction is examined in Sec.~\ref{sec:vertex} and Sec.~\ref{sec:energy}, respectively. A summary of the findings is provided in Sec.~\ref{sec:summary}. The complete detector survey results will be reported by the JUNO collaboration in the future, allowing for updates to this method as needed.

\section{Survey of the detector geometry during installation}\label{sec:survey}

\begin{figure}[!htb]   \includegraphics[width=\hsize]{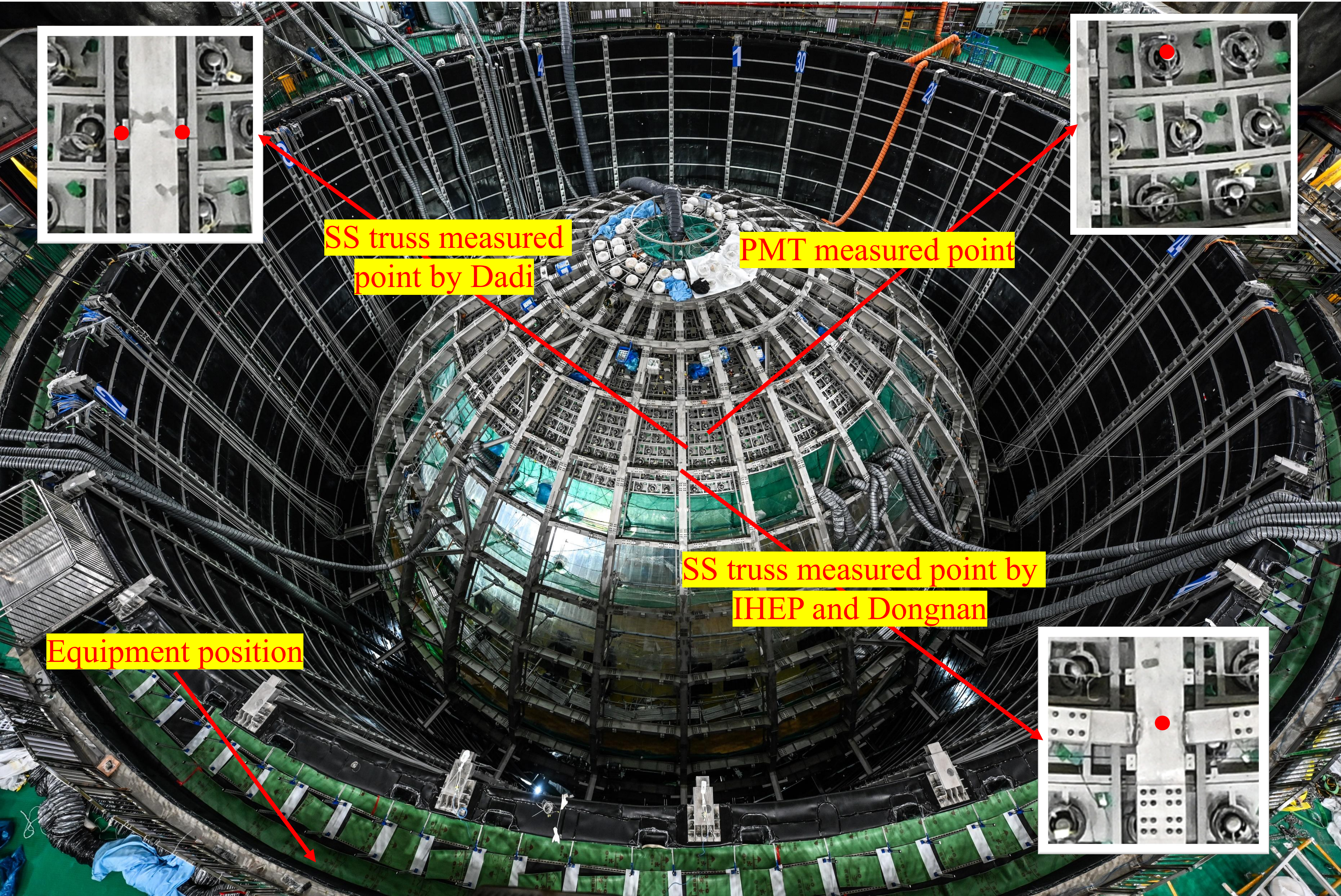}
\caption{Examples of measured positions of the SS shell and PMTs}
\label{fig:Measured_point}
\end{figure}

The primary configuration of the detector consists of a latticed shell made of stainless steel (SS) with an inner diameter of 40.1~m. It is structured with 23 latitudinal zones and 30 longitudinal zones (with 10 longitudinal zones at the top or bottom due to smaller radius). The shell is supported by 30 pairs of stainless steel legs in the lower hemisphere~\cite{JUNO:2023ete}. The placement of all the PMTs on the shell is facilitated by extended SS mounting modules, with each module capable of accommodating up to eight 20-inch PMTs fixed horizontally between two longitudinal beams. After construction was completed, surveys of the SS shell were conducted by two companies using Total Station, an electronic and optical instrument commonly employed in modern surveying with a precision of 2~mm. However, the bottom four latitudinal zones were finished two years later and were not included in these initial surveys. The SS shell construction company, denoted as Dongnan, measured the coordinates of 467 out of 650 intersections of longitudinal and latitudinal beams. 
An example measuring point is shown as the red dot in the bottom right zoom-in view in Fig.~\ref{fig:Measured_point}.
The PMT installation company, denoted as Dadi, measured the coordinates of 676 out of 1,630 fixing points for PMT modules on longitudinal beams. The top left zoom-in view in Fig.~\ref{fig:Measured_point}
shows an example measuring point of Dadi. After that, PMTs were installed from top to bottom on the detector. 
Surveys of the PMTs' position were conducted by the IHEP team in three batches over approximate one year during the installation. 
In this paper, only results from the first batch, where $z >$ 5~m, are analyzed. The complete geometry survey results will be reported by the JUNO collabortion in the future. Since PMTs are placed with the head facing inward, coordinates of the PMT tail were measured, as illustrated by the examplary red dot in the top right zoom-in view in Fig.~\ref{fig:Measured_point}.
A total of 808 out of 17,612 PMTs were measured together with 226 intersections of the SS shell. 
The IHEP surveys employed a Faro Vantage S laser tracker~\cite{laser_tracker} with 16~$\mu$m precision. However, considering potential equipment movement during measurements, the uncertainties for both the Total Station and laser tracker were estimated at 3~mm.  Measured positions of the SS shell and PMTs are shown in Fig.~\ref{fig:Measured_point} and summarized in Table~\ref{tab:PMT and SS truss condition}. 

\begin{table}[!htb]
\caption{Number of measured and total positions of PMTs and the stainless steel (SS) shell.}
\label{tab:PMT and SS truss condition}
\begin{tabular*}{\hsize}{@{\extracolsep{\fill}}lr}
\toprule
Items & Measured / Total  \\
\midrule
PMT & 808 / 17612 \\
SS shell (by IHEP) & 226 / 650 \\
SS shell (by Dongnan) & 467 / 650 \\
SS shell (by Dadi)  & 676 / 1630  \\
\bottomrule
\end{tabular*}
\end{table}

\subsection{Geometry survey of the SS shell}

\begin{figure}[!htb]
    \centering
    \includegraphics[width=\hsize]{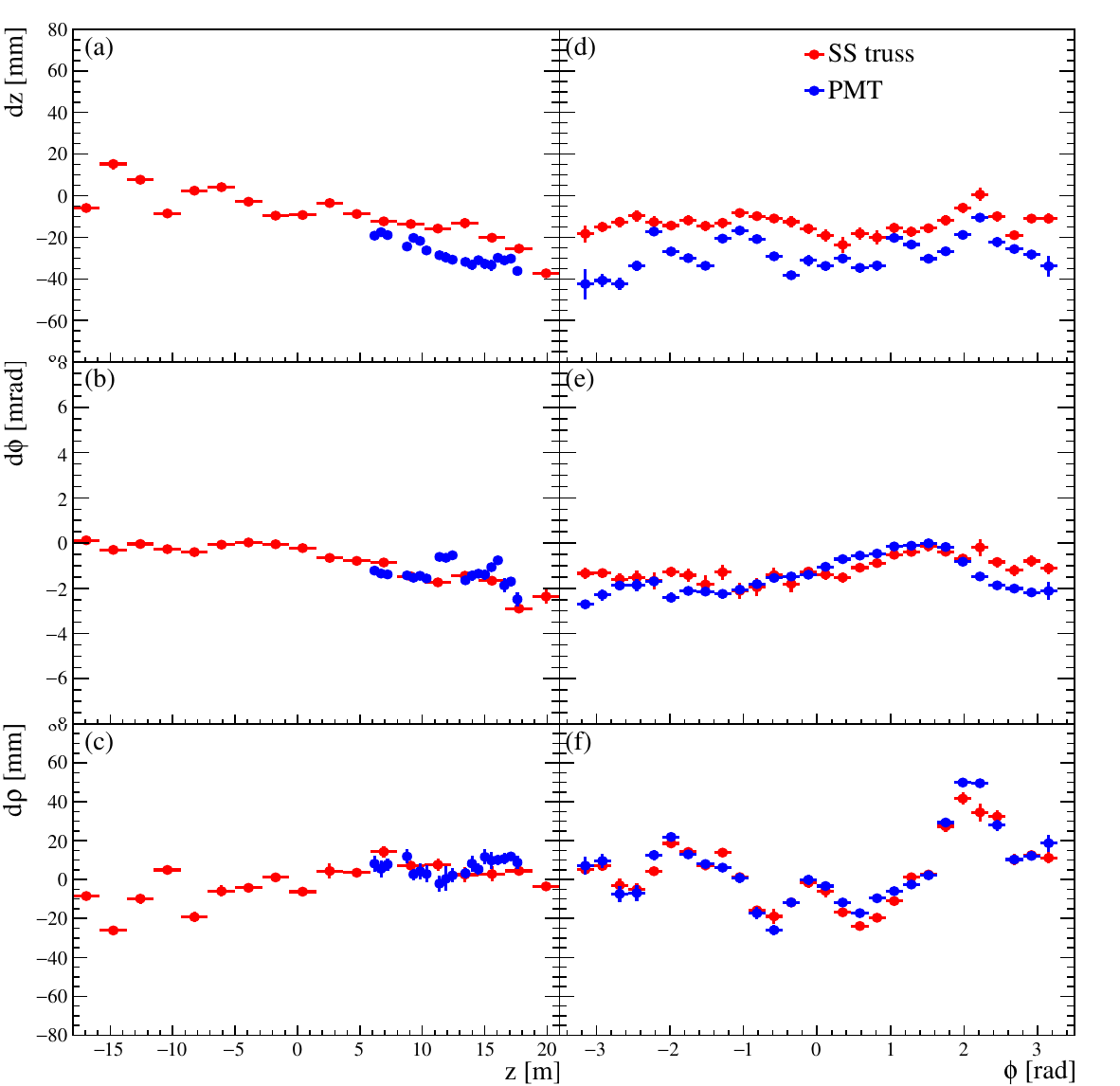}
    \caption{Comparison between surveyed and designed detector geometries, expressed in their differences d$z$, d$\phi$, and d$\rho$ as a function of $z$ in panels (a), (b), (c) and $\phi$ in panels (d), (e), (f). The stainless steel (SS) shell was measured by three groups Dongnan, Dadi, and IHEP, and their average values are used. PMTs were measured by IHEP. Note that panel (f) only includes data for $z > 5$~m to exclude averaging effects along the $z$-axis.}
    \label{fig:survey}
\end{figure}

All the measured results are represented in the cylindrical coordinate system, with the center of the water pool as the origin, which is also defined as the center of the CD. These results are compared with the designed values, and their differences are displayed in Fig.~\ref{fig:survey}, where $z$, $\phi$, and $\rho$ represent the height, azimuth angle, and the distance to the z-axis, respectively, within cylindrical coordinate system.  
For the SS shell, measurements from the three groups are averaged when they fall within the same bin and are represented by the red dots. 

In the lower hemisphere ($z<0$), the average deformation is within 1~cm in the $z$ and $\rho$ direction, and within 0.5~mrad (corresponding to 1~cm at the equator of the SS shell) in the $\phi$ direction. 
These results are expected because of the supporting legs, which provide significant constraints on the deformation. 
In the upper hemisphere ($z>0$), continuous drops are observed in panels (a) and (b), with a slope of d$z/z\approx$ 0.2~cm/m and d$\phi/z\approx$ 0.2~mrad/m, respectively.
These findings suggest slight sinking and twisting of the SS shell, presumably due to gravitational forces during installation, with the maximum deformation reaching approximately 3~cm near the top of the shell. 
Concerning deformation along the azimuth angle, there is no apparent dependence for d$z$ and d$\phi$ as depicted in panels (d) and (e). To mitigate the averaging effect induced by $\rho$-directional deformation across varying heights $z$, the data presented in panel (f) is restricted to $z > 5$~m, revealing a distinct variation of d$\rho$ with respect to $\phi$. The most significant deformation in the $\rho$ direction is approximately 4~cm at $\phi=2.2$~rad.

The deformation of the SS shell may vary under different loading conditions, such as the increased weight from the installed acrylic sphere and PMTs. To monitor this effect, three positions at a height of approximately +17~meters were randomly selected and measured three times during the installation to account for different load conditions. The results are summarized in Table~\ref{tab:loaded condition}. 
Compared to the initial measurement with no load, the SS shell descended by 3~mm when 386~tons of acrylic and PMTs were installed in the top hemisphere in June 2023. 
An additional 1~mm of deformation occurred three months later when the PMT installation progressed to the lower hemisphere, increasing the load to 575~tons. This deformation aligns with the Finite Element Analysis (FEA).
Further deformation is expected to be minimal due to additional supporting legs at the lower hemisphere. After  filling of liquid scintillator and water in and out of the acrylic sphere, approximately 3,000~tons of buoyancy will be expected, causing several millimeters of upward deformation according to the FEA.

\begin{table}[!htb]
\caption{Deformation of the SS truss in z-direction at +17~m under various loading conditions, obtained by averaging measurements from three locations.}
\label{tab:loaded condition}
\begin{tabular*}{\hsize}{@{\extracolsep{\fill}}lccc}
\toprule
Measured Time & 2022 & Jun. 2023 & Sep. 2023 \\
\midrule
Load (\si{\tonne}) & 0 & 386 & 575 \\
Deformation (\si{\mm}) & $-30.7\pm1.7$ & $-33.7\pm1.7$ & $-34.7\pm1.7$ \\
\bottomrule
\end{tabular*}
\end{table}

\subsection{Geometry survey of PMTs}

To allow for comparison with the designed PMT positions, the measured positions of the PMT tails ($\rho$, $\phi$, $z$) were translated into the centers of the PMT glass bulbs ($\rho'$, $\phi'$, $z'$), which are intended to be positioned on a sphere of diameter 19,434~mm. This conversion was achieved using Eq.~\ref{eq:conversion function},
\begin{eqnarray}
\label{eq:conversion function}
\begin{aligned}
\rho^{\prime} &=
\rho(1-\frac{H}{\sqrt{\rho^2+z^2}})\\
\phi^{\prime} &=
\phi\\
z^{\prime} &=
z(1-\frac{H}{\sqrt{\rho^2+z^2}})\\
H &= L-R
\end{aligned}
\end{eqnarray}
Here, $H$ is the distance from the center of the PMT glass bulb to its tail, which differs by design between different between MCP-PMTs and Hamamatsu PMTs. $L$ is the distance from the top of the PMT glass bulb to the tail. $R$ is 184~mm for MCP-PMTs and 190~mm for Hamamatsu PMTs provided by the manufacturers. In the laboratory, we measured $L$ of 15 MCP-PMTs and 14 Hamamatsu PMTs using a laser system. After subtracting the corresponding $R$, their average $H$ were determined to be 440.62 $\pm$ 0.33~mm and 514.81 $\pm$
0.57~mm, respectively. 

The differences between the converted PMTs positions and the designed values are also shown as the blue dots in Fig.~\ref{fig:survey}. For the region $z>$~5~m where both PMT measurements and SS truss data are available, the average differences are -10.0~mm, 0.4~mrad, and 2.9~mm in the $z$, $\phi$, and $\rho$ directions, respectively. On average, the PMTs' z-coordinate is approximately 1~cm lower than that of the SS shell, attributing to the gravitational forces acting on the PMTs and their SS modules. 
On the other hand, no systematic difference is observed between the PMTs and the SS shell in the $\phi$ and $\rho$ directions. 
The standard deviations in these directions are 5.0~mm, 0.6~mrad, and 5.4~mm, respectively, indicating a clear correlation between PMT positions variations and the SS shell deformation. 
As a result, we can utilize the collected PMT data to establish a model for predicting the positions of neighboring PMTs. By incorporating information obtained from the SS truss survey, this model can be extrapolated to areas where PMT surveys have not been conducted. The methodology for this modeling approach will be detailed in Sec.~\ref{sec:realization}.

\section{Modeling and implementation of the realistic detector geometry}
\label{sec:realization}
Based on the analysis and discussion in Sec.~\ref{sec:survey}, the average deformation of the SS shell in the lower hemisphere is less than 1~cm and there is no systematic deformation pattern. Therefore, PMTs in this region ($z<0$) are assumed to be at their designed locations. In contrast, the upper hemisphere exhibits systematic deformations in all directions, which must be accounted for in our analysis.

\subsection{Modeling of PMT positions}\label{sec:model}

\begin{figure}[!htb]  \includegraphics[width=\hsize]{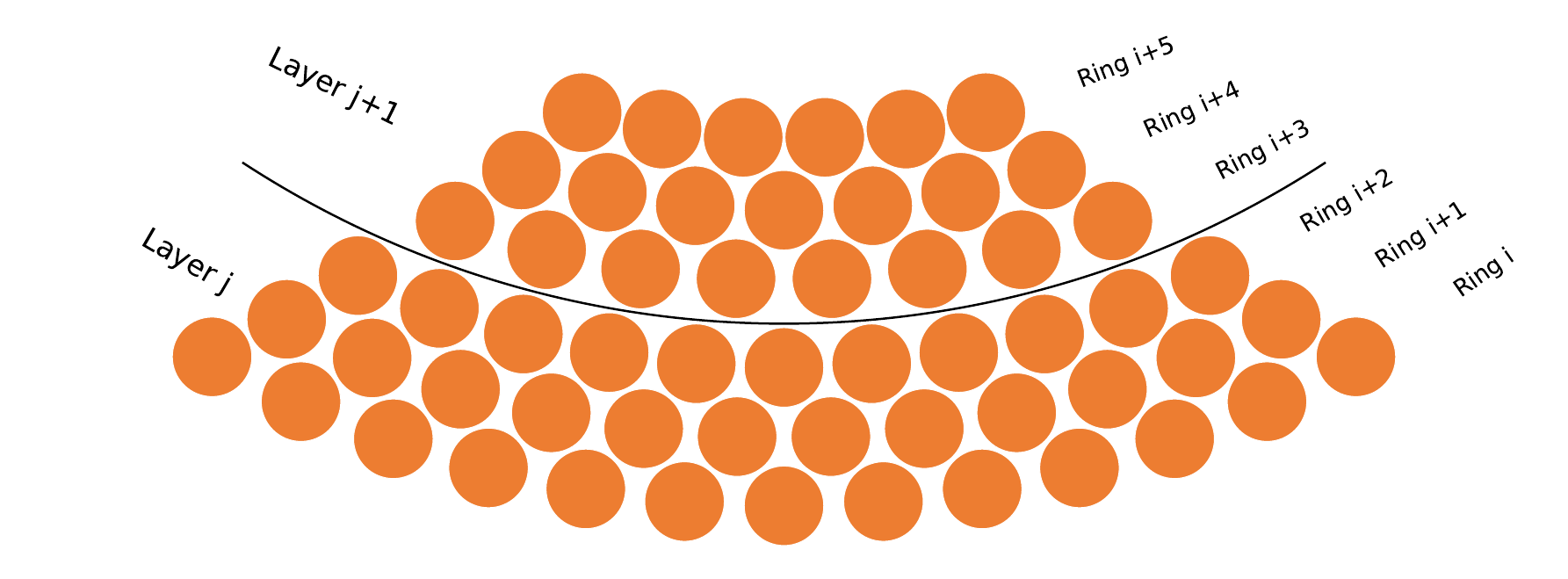}\caption{Demonstration of ring and layer.}
\label{fig:ring_layer}
\end{figure}

The specific positions of individual PMTs must be defined within the offline software to facilitate accurate detector simulation and event reconstruction. The modeling of PMT displacements is achieved by grouping the PMTs into rings and layers, as illustrated in Fig.~\ref{fig:ring_layer}. In the upper hemisphere, the PMTs are positioned along the $z$-axis in 62 concentric rings, from ring 1 located around the equator to ring 62 situated at the top of the detector. These rings are then categorized into 12 layers, designated as layer 1 through layer 12. The association between layer numbers and their corresponding ring numbers is outlined in Table~\ref{tab:Layers and rings}. Each layer, except for layer 12, accommodates a uniform maximum number of PMTs per ring, as detailed in Table~\ref{tab:Layers and rings}, with PMTs overlapping between neighboring rings. The actual quantity of installed PMTs in each ring is less due to design constraints caused by interference with the detector support structure. In layer 12, positioned at the top of the detector, PMTs cannot intersect due to spatial limitations, resulting in varying numbers of PMTs within each ring. The PMT configuration is symmetric in the lower hemisphere, thus employing a similar definition of rings and layers.

\begin{table}[!htb]
\caption{Definition of PMT ring numbers and layer numbers as well as their association in each of the detector hemisphere. Maximum number of PMTs installed in each ring is also introduced.}
\label{tab:Layers and rings}
\begin{tabular*}{\hsize}{@{\extracolsep{\fill}}lcccccc}
\toprule
Layer & 1 & 2 & 3 & 4 & 5 & 6 \\
\midrule
Ring & $1\sim13$ & $14\sim20$ & $21\sim25$ & $26\sim30$ & $31\sim34$ & $35\sim38$ \\
Max PMT & 219 & 205 & 191 & 174 & 159 & 142 \\
\midrule
Layer & 7 & 8 & 9 & 10 & 11 & 12 \\
\midrule
Ring & $39\sim41$ & $42\sim44$ & $45\sim46$ & $47\sim48$ & $49\sim50$ & $51\sim62$ \\
Max PMT & 128 & 114 & 104 & 90 & 82 & 71 \\
\bottomrule
\end{tabular*}
\end{table}

After introducing the rings and layers, the modeling of PMT displacements is constructed in the $z$, $\phi$, and $\rho$ directions, respectively. In this paper, we focus on the most significant dependencies: d$z$-$z$ (Fig.\ref{fig:survey}~(a)), d$\phi$-$z$ (Fig.\ref{fig:survey}~(b)), and d$\rho$-$\phi$ (Fig.\ref{fig:survey}~(f)) to demonstrate our modelling approach. However, this method can be extended to other directions or applied when additional survey data becomes available.

\begin{figure}[!htb]  \includegraphics[width=\hsize]{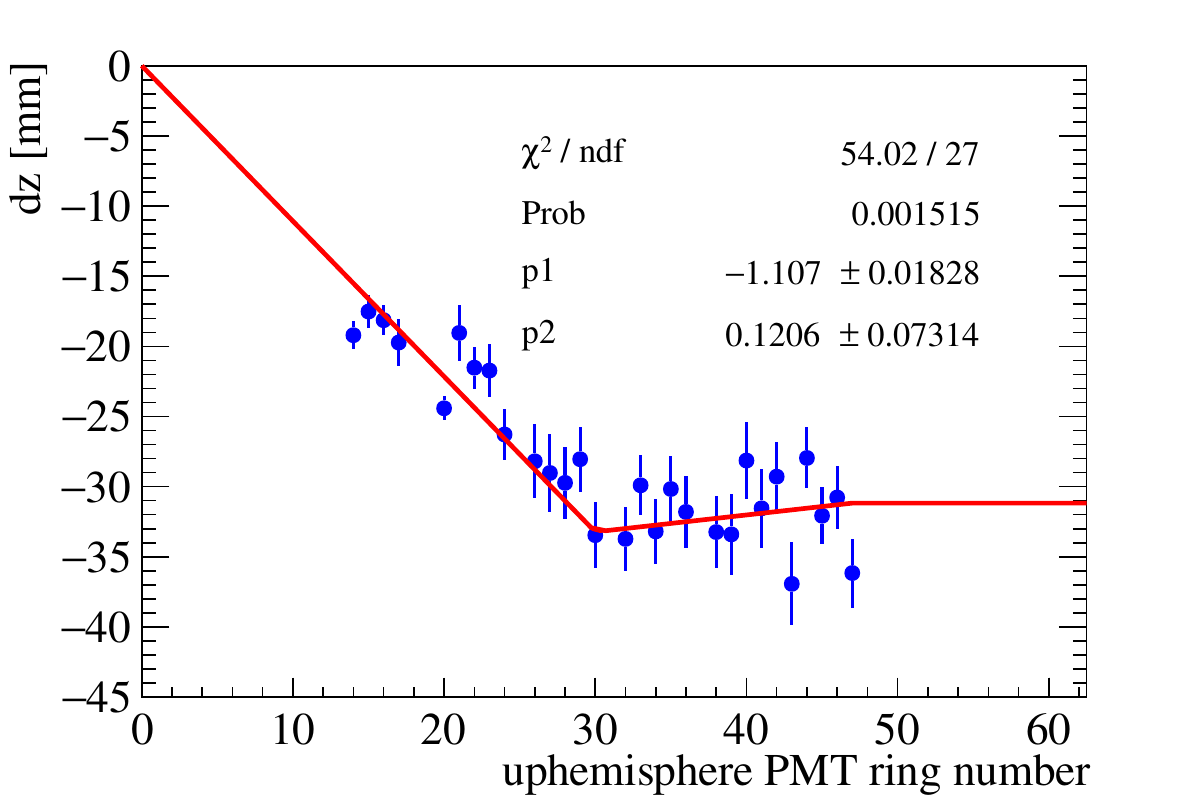}
\caption{Measured PMT's displacement in the $z$ direction as a function of ring number. The red line represents the best-fit piecewise function.}
\label{fig:modification of z}
\end{figure}

\paragraph{Modeling of PMT displacements in the $z$ direction}
The displacement of the measured PMTs in the $z$ direction is shown in Fig.~\ref{fig:modification of z} and can be characterized by a piecewise function presented in Eq.~\ref{eq:piecewise function}, with the intercept fixed at zero, as indicated by the SS shell deformation. A slight slope of the displacement was observed through the fitting parameter from ring~30 to ring~47, while PMTs beyond ring~47 assumed to exhibit a similar displacement as ring 47. 

\begin{eqnarray}
\label{eq:piecewise function}
    \begin{aligned}
f(z) =
\begin{cases}
p_{1}\cdot{\rm ring} & 0 \leq {\rm ring} < 30 \\
p_{2}\cdot({\rm ring} - 30) + 30~p_{1} & 30 \leq {\rm ring} < 47\\
17~p_{2} + 30~p_{1} & 47 \leq {\rm ring} \leq 62\\
\end{cases}
    \end{aligned}
\end{eqnarray}

\paragraph{Modeling of PMT displacements in the $\phi$ direction}
As explained in the beginning of this section, each layer consists of an equal number of PMTs per ring, with PMTs intersecting between adjacent rings. Therefore, no relative rotation in the $\phi$ direction is anticipated within a layer, although it may occur across different layers. Consequently, we determined the average d$\phi$ of the measured PMTs for each layer and then uniformly adjusted the orientation of all PMTs within the same layer by this angle d$\phi$. This procedure was applied to layers 2-10 where measurement data were available, while PMTs in other layers were assumed to remain in their original positions.

\begin{figure}[!htb]
\includegraphics[width=\hsize]{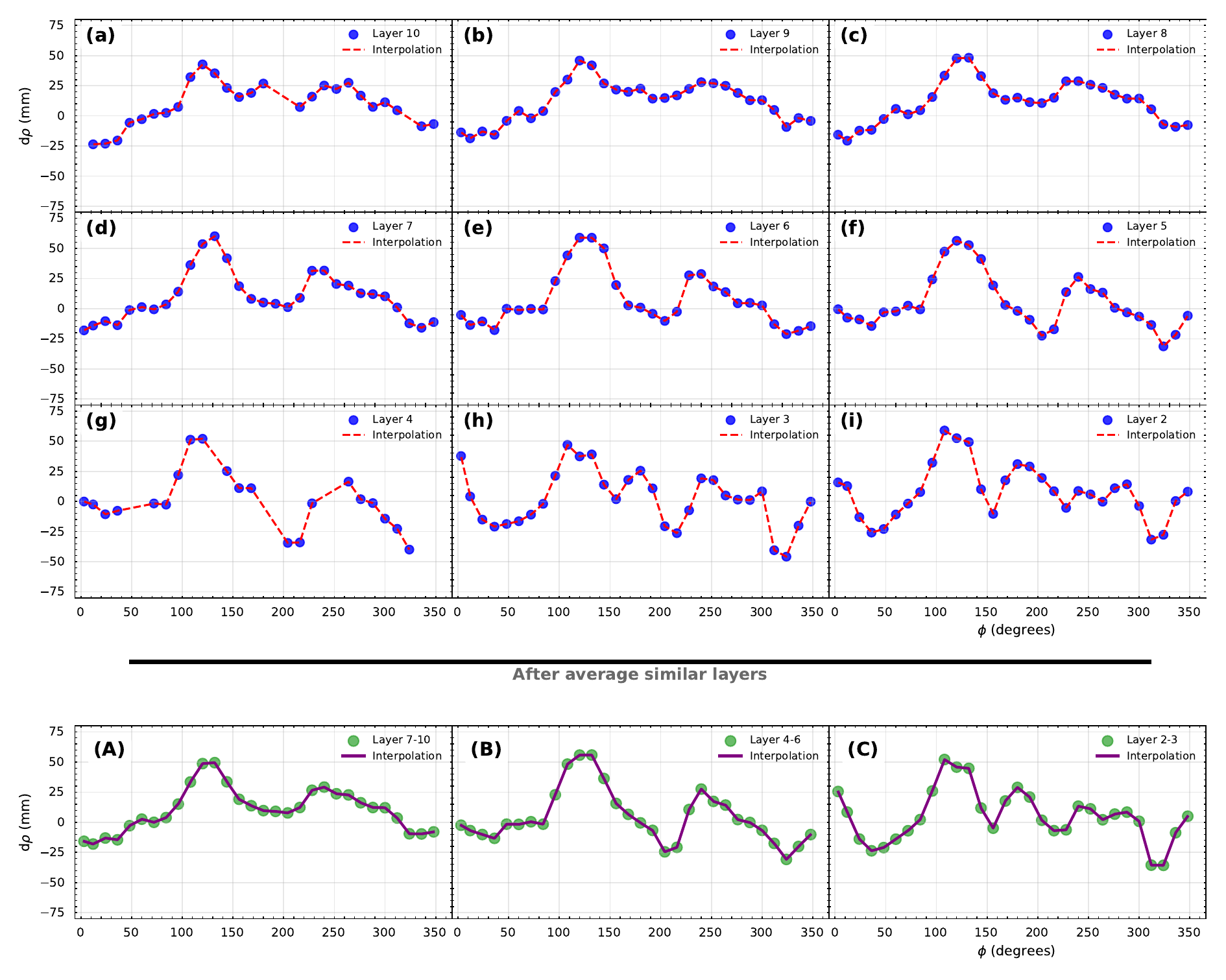}
\caption{(a)-(i), average displacement of PMTs in each model in $\rho$ direction. (A)-(C) Combined three sets of PMT displacement. Points are data and lines are linear interpolation.}
\label{fig:interpolation dr-phi}
\end{figure}

\begin{figure}[!htb]
\includegraphics[width=\hsize]{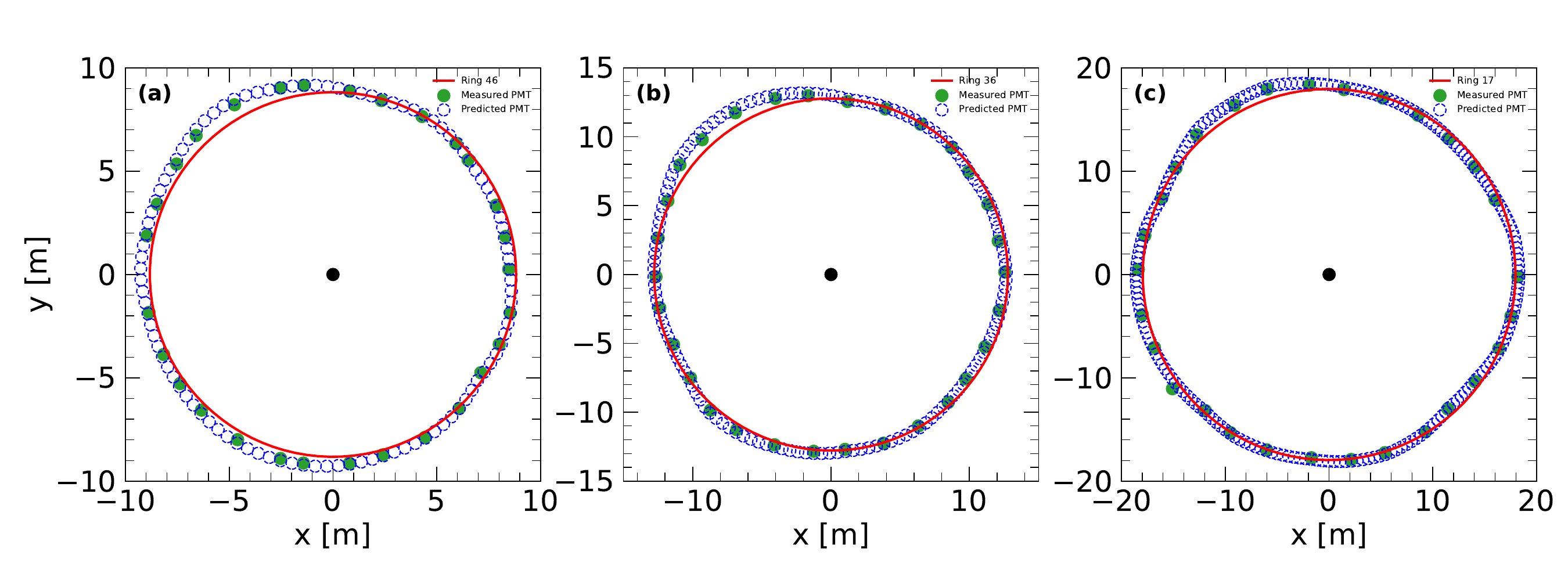}
\caption{Examples of PMTs' positions prediction in three rings. The red lines represent the ideal positions, while the green points and blue hollows indicate the measured and predicted positions, respectively. The actual displacement of the PMTs is on the order of centimeters, magnified by a factor of 20 in the figure for visualization.}
\label{fig:interpolation 3 rings}
\end{figure}

\paragraph{Modeling of PMT displacements in the $\rho$ direction}
Fig.~\ref{fig:interpolation dr-phi} shows the average displacement of PMTs in each model along the $\rho$ direction, d$\rho(\phi)$, as a function of azimuth angle $\phi$ for individual layers. Layers exhibiting similar PMT displacements are further combined to simplify the geometry model and mitigate the risk of geometry overlap in the detector simulation. Subsequently, three sets of d$\rho(\phi)$ data were acquired for layers 7-10, 4-6, and 2-3, respectively. Linear interpolation was then applied to predict each PMT's position, shown as the curves in the bottom three panels in Fig~\ref{fig:interpolation dr-phi}. Examples of PMTs' anticipated positions in three rings are displayed in Fig.~\ref{fig:interpolation 3 rings}, with the displacement magnified by a factor of 20 for enhanced clarity. PMTs in layers 1, 11, and 12 are presumed to remain stationary.

\subsection{Implementation and validation of the new geometry}

In the JUNO simulation software framework~\cite{Lin:2022htc}, a new geometry was implemented to assess its effect on the detector performance, with PMTs' positions adjusted using the model introduced in Sec.~\ref{sec:model}. Considering the minimum clearance between two adjacent PMTs is 3~mm while the PMT displacement is up to a few centimeters, the primary challenge in implementing the realistic geometry lies in preventing PMT overlaps. Within the Geant4-based simulation software, there exists a functionality to detect overlaps and report the count of PMTs involved in such instances. The geometry adjustments were carried out sequentially in one direction at a time, followed by a thorough examination for PMT overlaps. The outcomes of this analysis are summarized in Table~\ref{tab:differnt direction overlap}.

\begin{table}[!htb]
\caption{Comparison of the overlapping PMT numbers for different geometry configurations}
\label{tab:differnt direction overlap}
\begin{tabular}{lr}
\toprule
Geometry & Overlap number \\
\midrule
ideal geometry & 16 \\
modification $z$ & 16 \\
modification $z + \phi$ & 25 \\
modification $z + \phi + \rho$ & 93 \\
\bottomrule
\end{tabular}
\end{table}

The default detector geometry with the ideal PMT positions has 16 instances of overlaps. This number remains consistent when solely modifying the $z$-coordinate. With adjusting both $z$ and $\phi$ coordinates, the count of overlaps rises to 25, demonstrating a minor increase attributed to the $\phi$ modification. Subsequently, upon introducing modifications to the $\rho$-coordinate in addition to $z$ and $\phi$ the count increases to 93 overlaps. This substantial change can be attributed to the non-systematic variations in $d\rho$ as a function of $\phi$, coupled with the 3~mm spacing between adjacent PMTs. Notably, the JUNO CD houses 17,612 20-inch PMTs, resulting in an approximate 0.5\% overlap rate and significantly lower overlap volumes, rendering its impact negligible. Consequently, the revised geometry were used for subsequent analyses.

\begin{figure}
\includegraphics[width=\hsize]{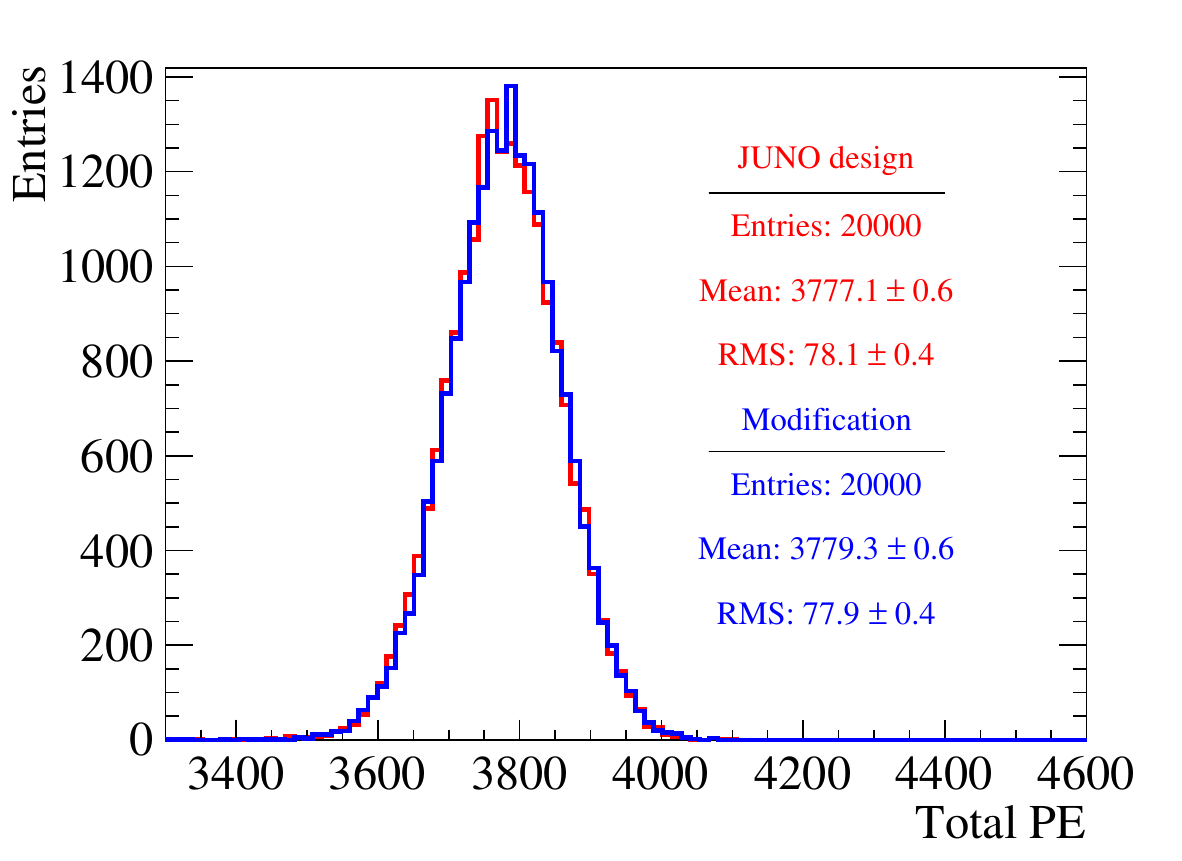}
\caption{Simulated number of photoelectrons in case of designed geometry (red) and modified geometry (blue).}
\label{fig:light yield}
\end{figure}

Additionally, in JUNO simulation, the PMTs' module is substituted with a reflective surface positioned in close proximity to the PMT tail. When the cylindrical coordinate $\rho$ is adjusted based on d$\rho(\phi)$, the reflective surface must be extended outward accordingly to avoid overlapping with the PMTs. To simply the simulation of the revised geometry, the spherical radius of the reflective surface is enlarged by 45~mm, slightly exceeding the maximum spherical radius increase of the PMTs. For consistency and to ensure a fair comparison between the original and revised geometries, the spherical radius of the reflective surface is augmented by the same increment in the original design. This approach enables a focused analysis of the impact of the new geometry. Subsequently, 20~k AmC events at the detector center were simulated using either the original or revised geometries. The light yield was compared by examining the n-H capture events in both datasets. Fig.~\ref{fig:light yield} illustrates the comparison of the total number of detected photons across all PMTs. The light yield is defined as below:
\begin{eqnarray}
\label{eq:light yield}
    \mathrm{Light   ~Yield} = \frac{\mathrm{Total~PE}}{2.22~\mathrm{MeV}}
\end{eqnarray}
where Total PE is the average value of total number of photoelectrons, and 2.22~MeV is the energy of the gamma emitted by neutron capture on hydrogen. 

Meanwhile one can also directly estimate the coverage of PMTs as below:
\begin{eqnarray}
\label{eq:coverage}
  \mathrm{coverage} \approx \sum_{i=1}^{17612}\frac{\pi r_{\mathrm{PMT}}^{2}}{4\pi d_{i}^{2}}
\end{eqnarray}
where $r_{\mathrm{PMT}}$ represents the radius of the PMT photon cathode and $d_{i}$ donate the distance between the $i$~th PMT and the center of the detector, calculated based on the PMT coordinates. In the designed geometry, all PMTs have the same distance $d_{i}$ of 19,434~mm, whereas in the new geometry, $d_{i}$ could vary from one PMT to another. A comparison of the light yield and PMT coverage between the two geometries is summarized in Table~\ref{tab:light yield and coverage talbe}. It is evident that the rise in light yield with the new geometry aligns closely with the PMT coverage increase. Furthermore, the light yield shows a marginal increase of only 0.059\%, suggesting that the alterations in geometry have a minimal effect on the light yield.

\begin{table}[!htb]
\caption{Comparison of the light yield and PMT coverage between the ideal geometry and measured geometry.}
\label{tab:light yield and coverage talbe}
\begin{tabular*}{\hsize}{@{\extracolsep{\fill}}l *{3}{c}}
\toprule
& \raisebox{-0.55\normalbaselineskip}{\begin{tabular}{@{}c@{}}Designed \\ Geometry\end{tabular}}
& \raisebox{-0.55\normalbaselineskip}{\begin{tabular}{@{}c@{}}Measured \\ Geometry\end{tabular}}
& \raisebox{-0.55\normalbaselineskip}{\begin{tabular}{@{}c@{}}Relative \\ Ratio\end{tabular}} \\
\midrule
PMT Coverage & 75.213\% & 75.254\% & +0.055\% \\
Light Yield (\si{\MeV}) & 1701.4 & 1702.4 & +0.059\% \\
\bottomrule
\end{tabular*}
\end{table}

\section{The impact of new geometry on event reconstruction}
The primary goal of JUNO is to determine the neutrino mass ordering. One of the main challenges is the unprecedented energy resolution of 3\% at 1~MeV. 
Given the importance of PMTs positions in the event reconstruction, one needs to study the impact of deformation of the CD on the event reconstruction performance, particularly the energy resolution.
In order to address this question, the following three cases listed in Table~\ref{tab:reconstruction data setting}  were studied.

\begin{table}[!htb]
\caption{Reconstruction data setting.}
\label{tab:reconstruction data setting}
\begin{tabular*}{\hsize}{@{\extracolsep{\fill}}cccc}
\toprule
 Case & Simulation Geometry & Reconstruction Geometry & Color\\
\midrule
1 & Designed & Designed & Blue\\
2 & Measured & Designed & Black \\
3 & Measured & Measured & Red\\
\bottomrule
\end{tabular*}
\end{table}

Case 1 represents the ideal case without any detector deformation. In this case the designed geometry is used for both the event simulation and reconstruction. It also serves as the reference case for comparison.
In Case 2, the events are simulated with the measured geometry, but during the reconstruction the designed geometry is used. This case is also referred to as "wrong geometry". It showcases what would happen if the geometry is not updated for the real data.
Case 3 is referred to as the "realistic geometry", it represents the realistic case in which the measured geometry is used for both the simulation and reconstruction. 

Two different sets of Monte Carlo samples were produced with the designed geometry and measured geometry, respectively, using JUNO simulation software~\cite{Lin:2022htc,Li:2017zku}. The details are summarized in Table~\ref{tab:MC sample}. 
Positron events with different kinetic energy were produced and they were distributed uniformly in the CD. They were used to evaluate the reconstruction performance. Meanwhile AmC events were produced at the CD center and used to define the energy scale.
$^{68}\mathrm{Ge}$ events with various positions following the recommendation in Ref.~\cite{Huang:2021baf} were produced to extract the expected time and charge response of PMTs, which are the crucial inputs to the event reconstruction described below.

\begin{table}[!htb]
\caption{List of the MC simulation samples: $e^{+}$ is the event we need to reconstruct and AmC is used to scale the energy}
\label{tab:MC sample}
\begin{tabular*}{\hsize}{@{\extracolsep{\fill}}l c c l}
\toprule
Source & Energy & Statistics & Position \\
\midrule
$e^{+}$ 
& \begin{tabular}[c]{@{}c@{}}$E_{k} = (0,0.5,1,2,3,$ \\ $4,5,8)$ \si{\MeV}\end{tabular}
& 50 k/\si{\MeV} 
& uniform in CD \\
AmC & 2.22 \si{\MeV} & 20 k & center of CD \\
$^{68}\mathrm{Ge}$ & 1.022 \si{\MeV} & 20 k/point & 293 points \\
\bottomrule
\end{tabular*}
\end{table}

The event reconstruction method is taken from Ref.~\cite{Huang:2022zum}.
It is a data driven simultaneous vertex and energy reconstruction algorithm. It utilizes calibration events at different positions to obtain the expected charge and time response of PMTs, both of which highly depend on the coordinates of each PMT, as well as the vertex and energy of the event. 
By comparing the observed and expected charge and time information of PMTs, a likelihood function was constructed and subsequently minimized to simultaneously reconstruct the vertex and energy.
Based on the description above, one can see that accurate PMTs coordinates must be used in extracting the expected charge and time response as well as in the event reconstruction.
Using the configurations in Table~\ref{tab:reconstruction data setting} and the samples in Table~\ref{tab:MC sample}, the impact of the new geometry on the event reconstruction is studied in details in the following subsections.

\subsection{Vertex reconstruction}
\label{sec:vertex}
\begin{figure}[!htb]
    \subfigure[]{
    \label{fig:vertex_z_a}
    \includegraphics[width=\hsize]{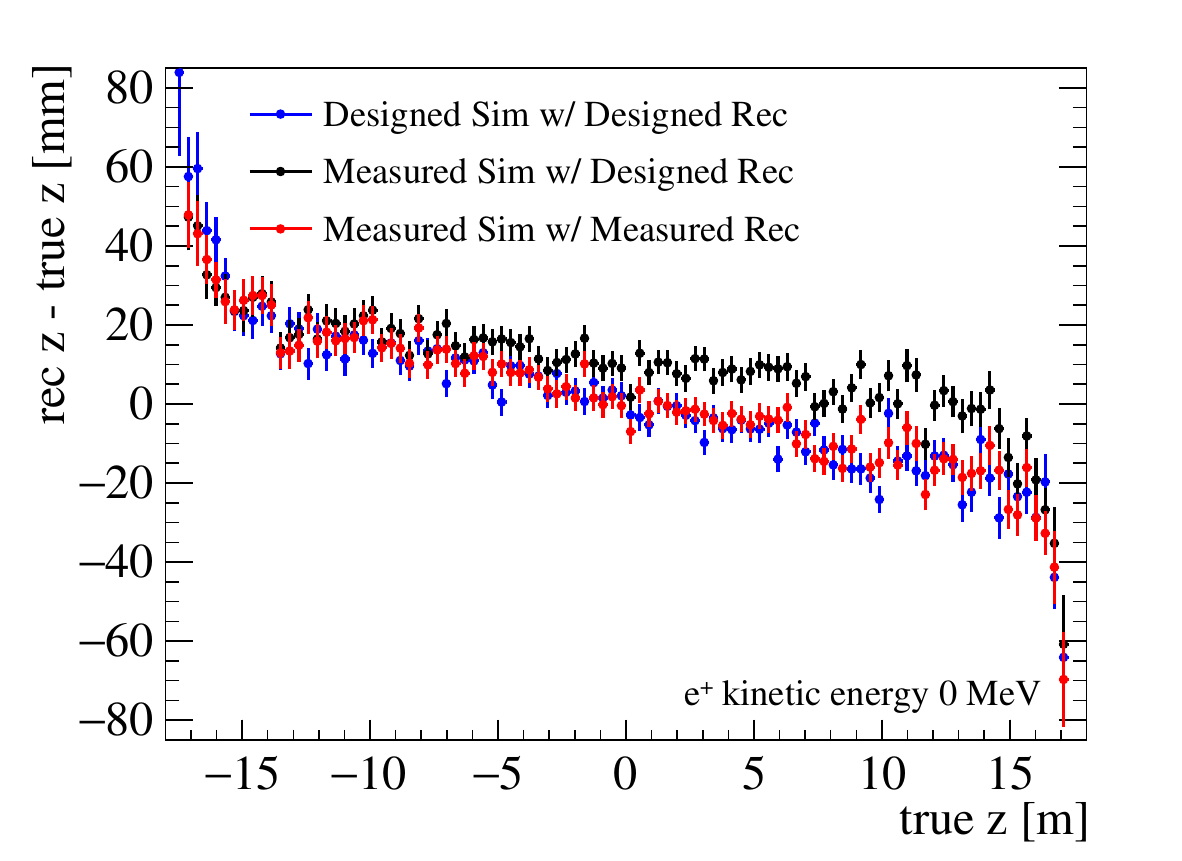}
    }
    \subfigure[]{
    \label{fig:vertex_z_b}
    \includegraphics[width=\hsize]{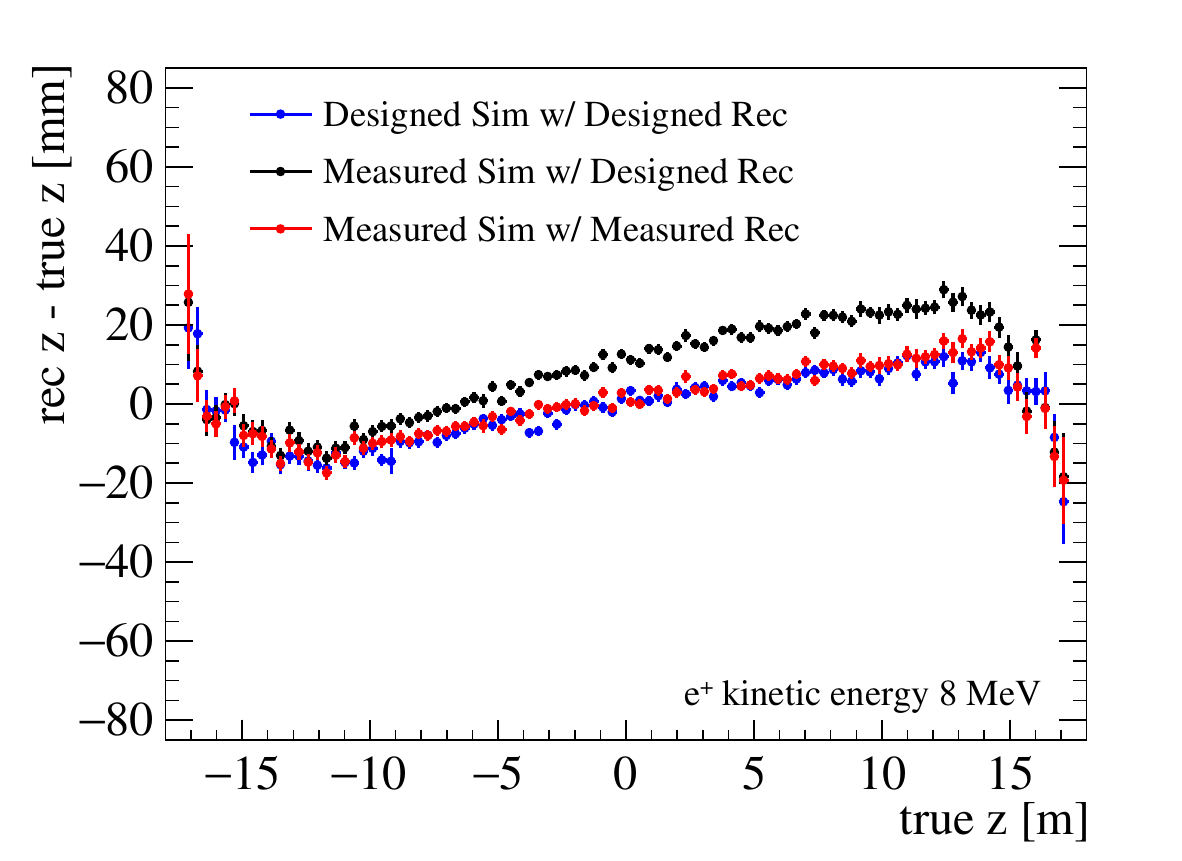}
    }
\caption{Vertex reconstruction bias in the $z$-direction versus $z$ for $e^{+}$ events with kinetic energies of (a) 0~MeV and (b) 8~MeV.}
\label{fig:vertex_z}
\end{figure}

Given that the new geometry is modeled using the survey data under the cylindrical coordinate system, we will evaluate the performance of the vertex reconstruction using the same $z$, $\rho$ and $\phi$ coordinates.
For the real data, the PMT coordinates have changed with respect to the designed values due to detector deformation. One should use the measured coordinates accordingly in the event reconstruction. 
Fig.~\ref{fig:vertex_z} shows the comparison of the $z$ reconstruction results for the three cases in Table~\ref{tab:reconstruction data setting}.  The upper and lower plots correspond to uniformly distributed positron events with 0 or 8~MeV kinetic energy respectively.  
One can see that Case 3 (red dots) which represents the realistic case yields almost the same performance as the ideal case (blue dots) without detector deformation. 
This indicates that once the realistic geometry is accurately modeled and used in reconstruction, the deformation has almost negligible impact on the $z$ reconstruction.
However, if the PMT coordinates are not updated and the designed values are still used for real data, this inconsistency would have a large impact on the reconstruction.
Since the PMT $z$ coordinates are shifted downward with respect to the designed values,  the reconstructed $z$ in Case 2 was pulled up by a maximum of about 20~mm, as shown by the black dots of Fig.~\ref{fig:vertex_z}.  

\begin{figure}[!htb]
    \subfigure[]{
    \label{fig:vertex r a}
    \includegraphics[width=\hsize]{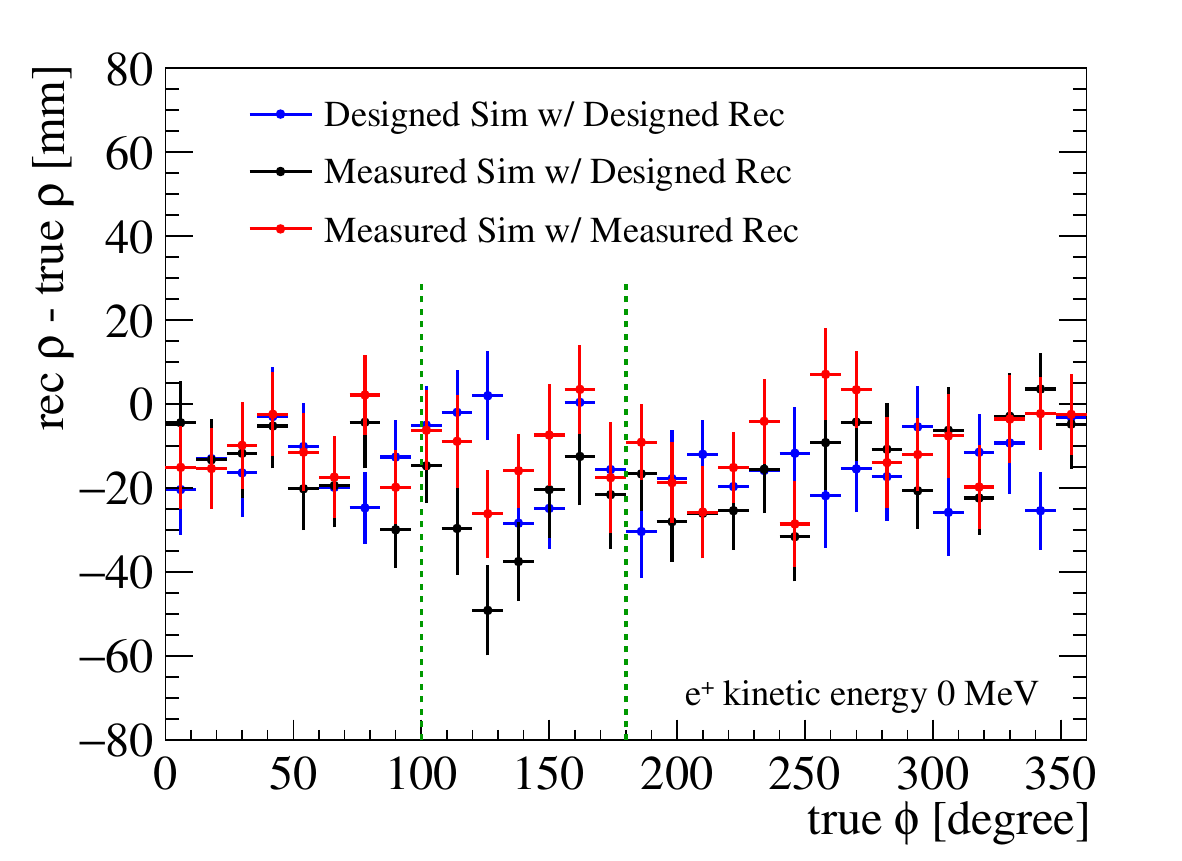}
    }
    \subfigure[]{
    \label{fig:vertex r b}
    \includegraphics[width=\hsize]{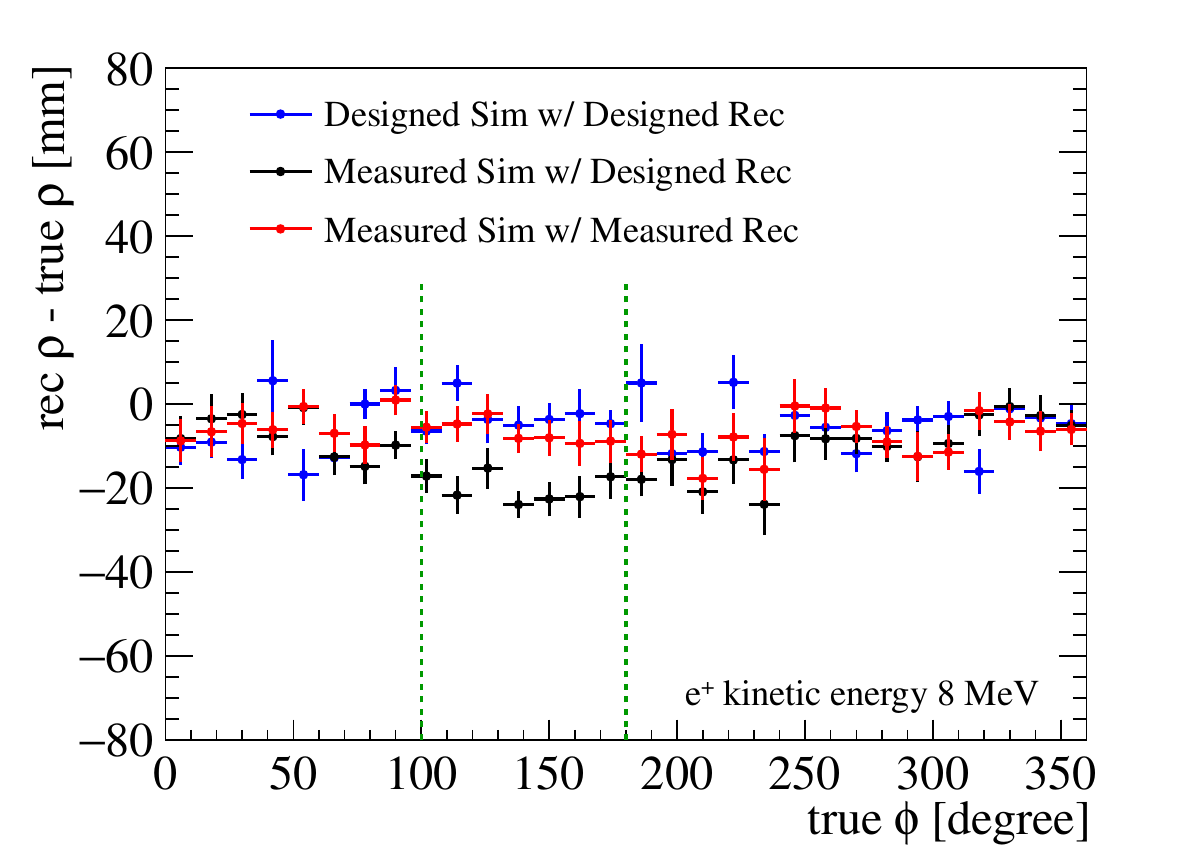}
    }
\caption{Vertex reconstruction bias in the $\rho$-direction versus $\phi$ for $e^{+}$ events with kinetic energies of (a) 0~MeV and (b) 8~MeV.}
\label{fig:vertex r}
\end{figure}

For the $\rho$ coordinate, the comparison of its reconstruction results were shown in Fig.~\ref{fig:vertex r}. Similar to $z$, Case 1 and Case 3 yield similar results, indicating that the impact of the detector deformation could be largely mitigated as long as the correct geometry is used. 
Meanwhile as shown by the black dots in Case 2, using inconsistent geometry has a non-negligible impact, particularly near the region around $120^\circ$ as indicated between the two vertical dashlines. 
The reconstructed $\rho$ was pulled towards the detector center by up to approximately 40~mm, since it has a outward shift in the realistic geometry with respect to the designed geometry, as one can see exactly from the left and middle plots in Fig.~\ref{fig:interpolation 3 rings}. Since the deformation in the $\rho$ direction differs between layers, to avoid the effects of averaging, we selected only events with vertices in the region 13~m $< z <$ 16~m in order to clearly demonstrate the impact of the $\rho$ direction deformation on vertex reconstruction.

\begin{figure}[!htb]
    \subfigure[]{
    \label{fig:vertex phi a}
    \includegraphics[width=\hsize]{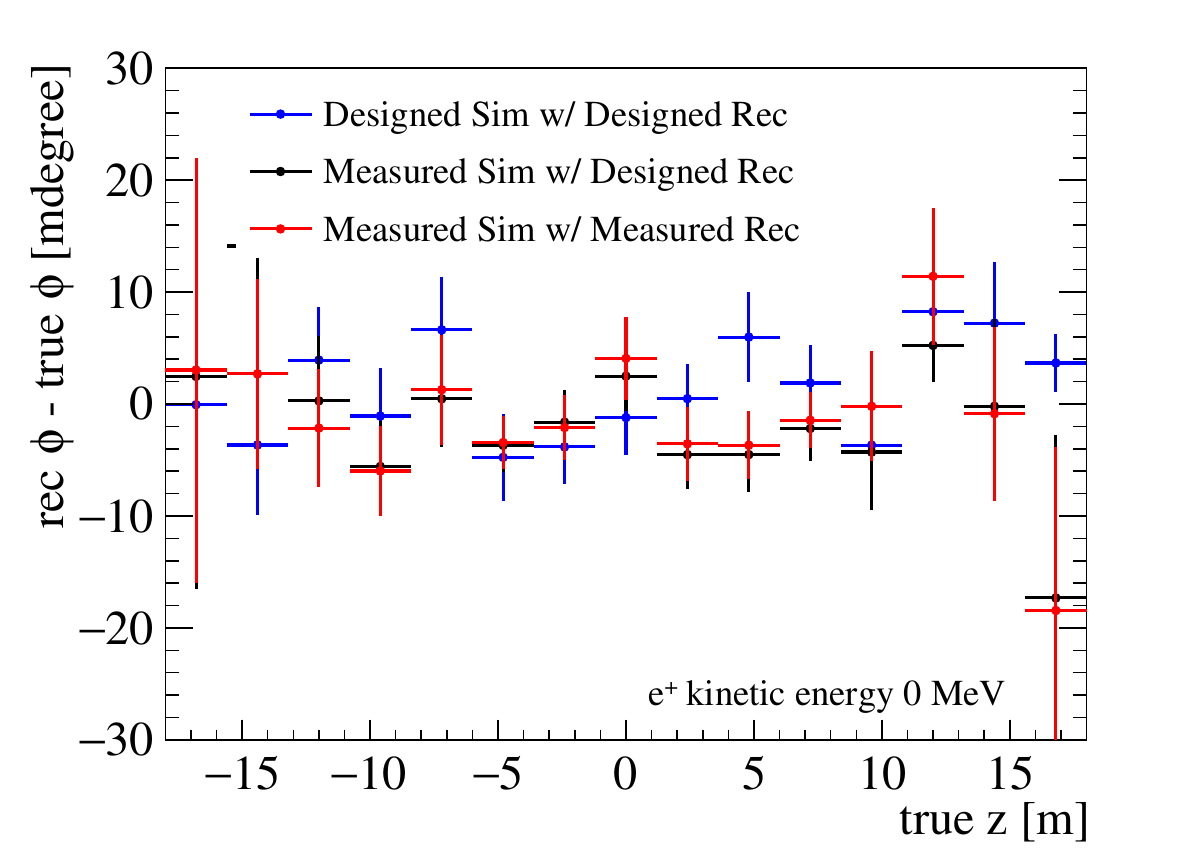}
    }
    \subfigure[]{
    \label{fig:vertex phi b}
    \includegraphics[width=\hsize]{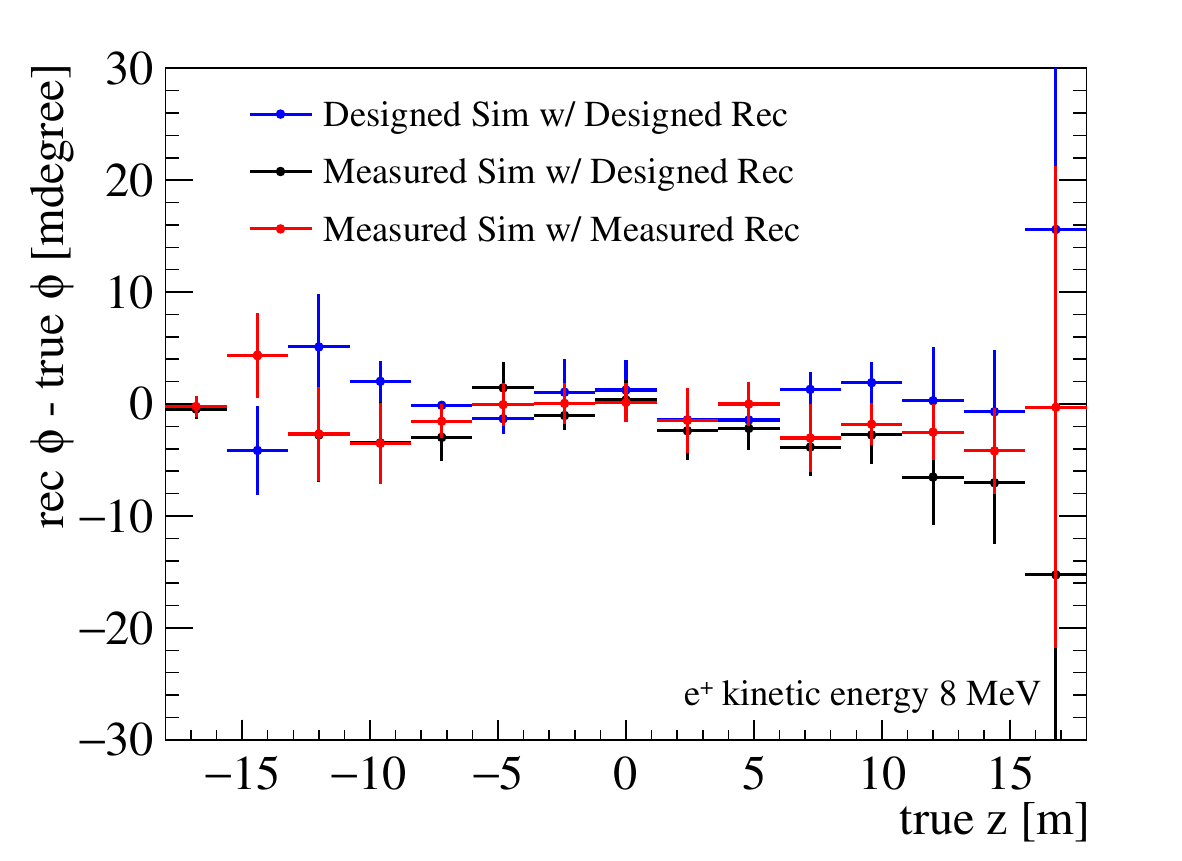}
    }
\caption{Vertex reconstruction bias in the $\phi$ direction versus $\phi$ for $e^{+}$ events with kinetic energies of (a) 0~MeV and (b) 8~MeV.}
\label{fig:vertex phi}
\end{figure}

For the $\phi$ coordinate, as Fig.~\ref{fig:survey}~(c) shows $d\phi$ is very small in the new geometry, the reconstruction algorithm is not sensitive to this small change in $\phi$. Thus no significant change of reconstructed $\phi$  was observed among the three cases, 
as shown in Fig.~\ref{fig:vertex phi}. 

\subsection{Energy reconstruction}
\label{sec:energy}

Similar to Sec.~\ref{sec:vertex}, the three cases from Table~\ref{tab:reconstruction data setting} were studied to check the impact of the new geometry on energy reconstruction. The performance of the energy reconstruction was evaluated in terms of energy resolution, energy non-linearity, the energy non-uniformity.

\begin{figure}[!htb]
    \subfigure[]{
    \label{fig:Eres_NL_a}
    \includegraphics[width=\hsize]{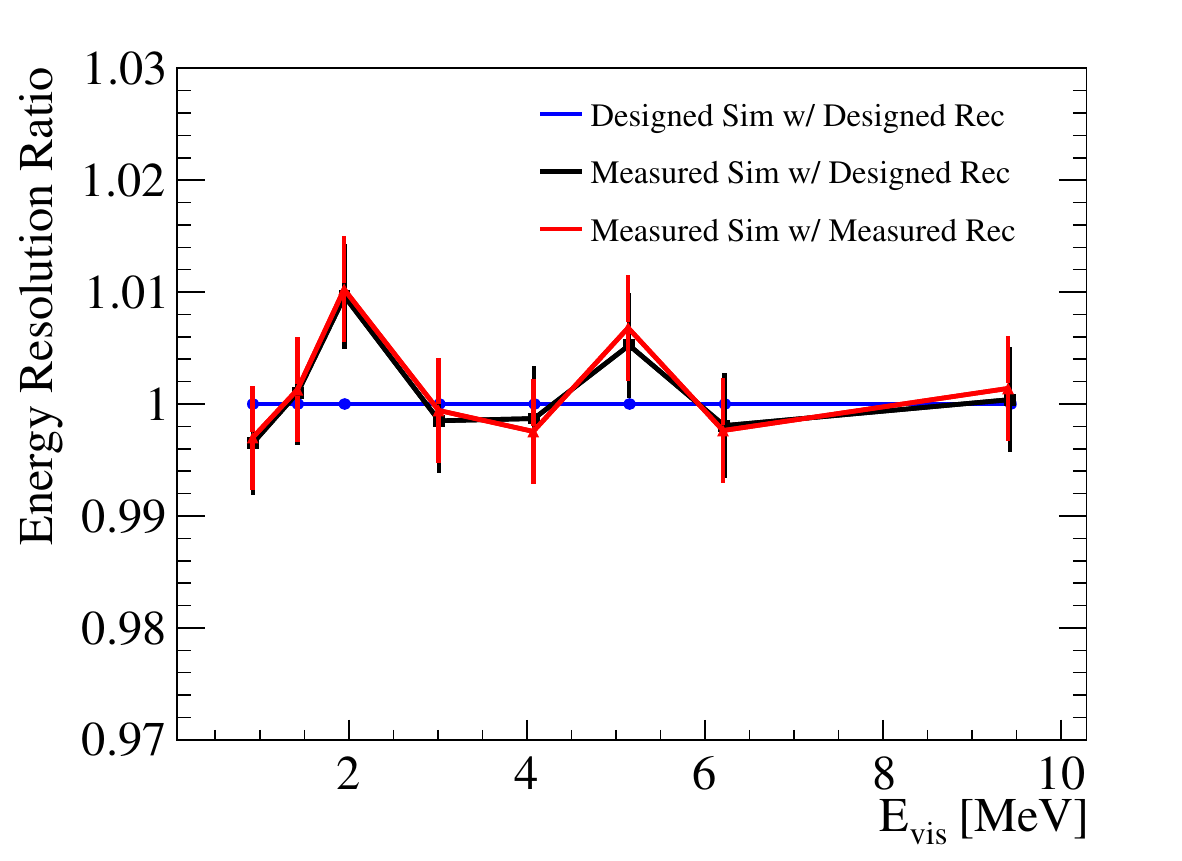}
    }
    \subfigure[]{
    \label{fig:Eres_NL_b}
    \includegraphics[width=\hsize]{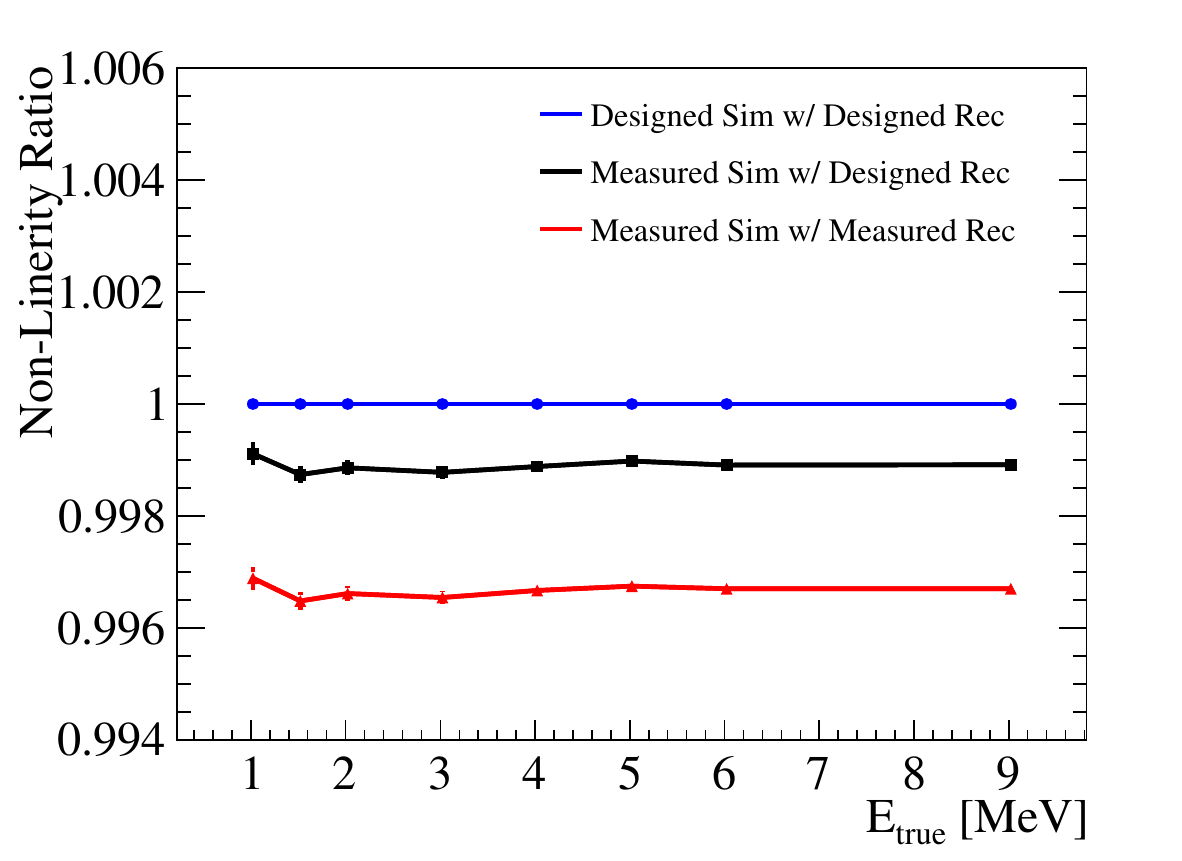}
    }
\caption{Energy reconstruction results for the three datasets: (a) shows energy resolution, and (b) shows the non-linearity.  The blue, black, and red curves represent Case 1, 2, and 3, respectively, as defined in Table \ref{tab:reconstruction data setting}.}
\label{fig:Eres_NL}
\end{figure}

Fig.~\ref{fig:Eres_NL}~(a) compares the energy resolution across the three cases. 
The blue curve represents Case 1, which serves as the reference and therefore appears flat. The black and red curves show the ratios of Case 2 and Case 3 to Case 1, respectively. When switching to the simulation using measured geometry, there is a random fluctuation below 1\%. The choice between measured or designed geometry in the reconstruction has negligible impact.  Similar comparison of energy non-linearity is presented in Fig.~\ref{fig:Eres_NL}~(b), where Case 1 is also defined as the reference. The impact of the new geometry is found to be less that 0.4\%.

\begin{figure}[!htb]
    \subfigure[]{
    \label{fig:compare Energy non-uniformity a}
    \includegraphics[width=\hsize]{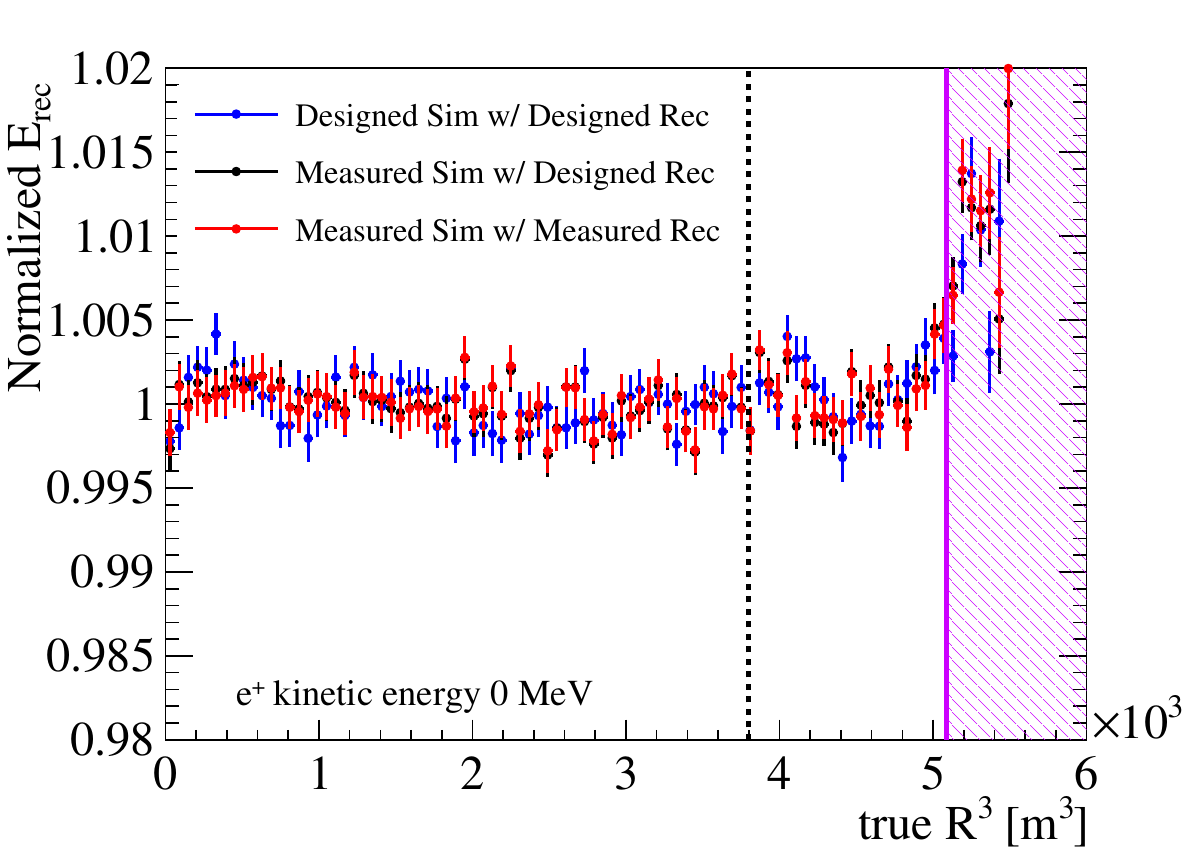}
    }
    \subfigure[]{
    \label{fig:compare Energy non-uniformity b}
    \includegraphics[width=\hsize]{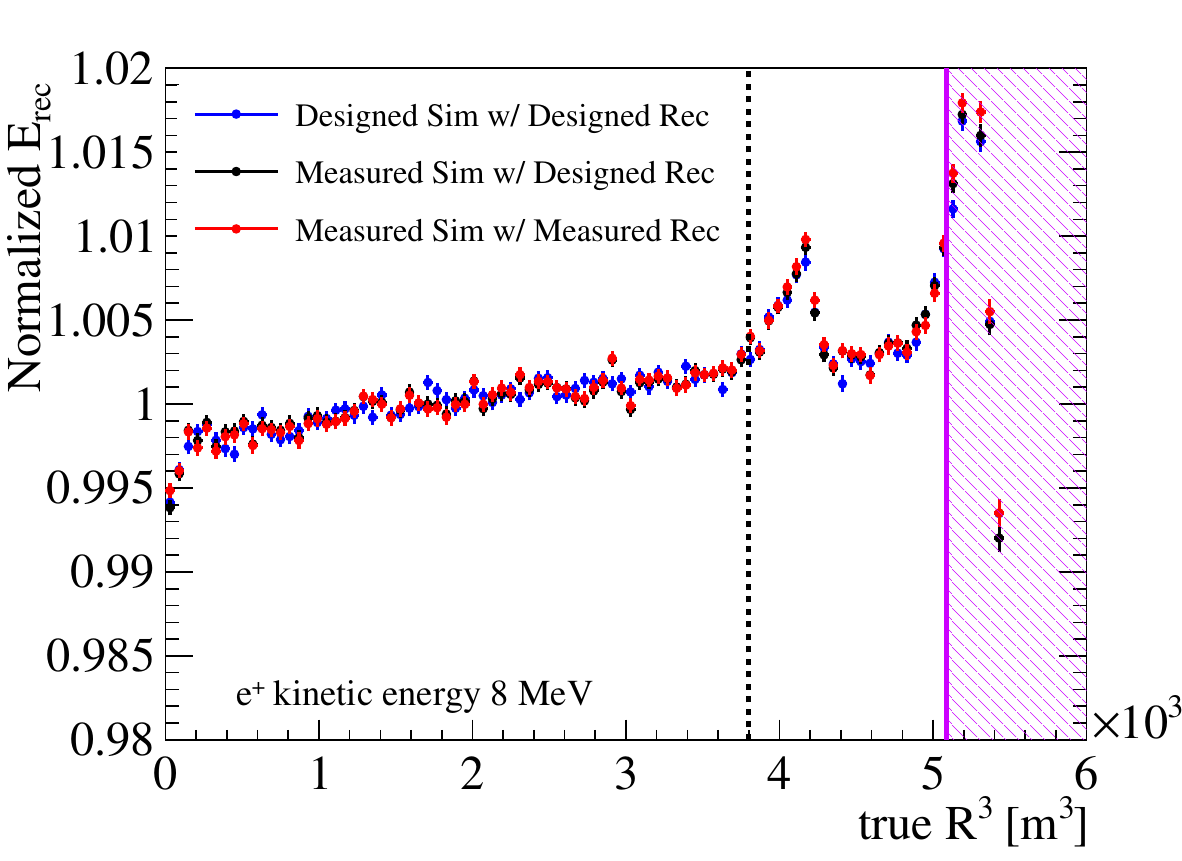}
    }
\caption{Energy non-uniformity versus $R^{3}$ for $e^{+}$ events with kinetic energies of (a) 0~MeV and (b) 8~MeV. The total reflection region and the fiducial volume cut are indicated by the dashed line and the shaded area, respectively.}
\label{fig:compare Energy non-uniformity}
\end{figure}

The comparison of the energy non-uniformity is shown in Fig.~\ref{fig:compare Energy non-uniformity}. 
Plots (a) and (b) correspond to the results of position events with 0 and 8~MeV kinetic energy, respectively. The results for all three cases are similar, and this pattern holds true for other energy points as well. These findings indicate that the new geometry has virtually no impact on energy non-uniformity. 

As described by sec.~\ref{sec:vertex} and sec.~\ref{sec:energy}, the impact of the new geometry on the event reconstruction was validated by comparing the three cases. The impact of detector deformation on the vertex reconstruction could be largely mitigated if the measured geometry were used during the reconstruction. Otherwise the inconsistent geometry would lead to bias for both $z$ and $\rho$ coordinate. Meanwhile the impact of detector deformation on the energy reconstruction is small, less than 1\% and 0.4\% on the energy resolution and non-linearity respectively.

\section{Summary and pespective}
\label{sec:summary}

In this paper, we first present the geometric survey results of the SS truss and a subset of the installed PMTs. A significant association has been identified between the SS truss and the PMTs' locations. Subsequently, a predictive model has been developed to estimate the placement of individual PMTs by interpolating the positions of the measured PMTs. This model has been extended to areas where data is solely available from the survey of the SS truss. The positions of each PMT were adjusted in the detector simulation based on the predictions, and the distribution of collected photoelectrons was examined, showing agreement with the expectation. The impact on the performance of event reconstruction was investigated by varying the geometry configuration in the simulation and reconstruction. Assuming the accuracy of the detector geometry in simulation, regardless of whether the geometry in reconstruction is accurate or not, the energy resolution was observed to remain consistent within a statistical uncertainty of less than 1\%. However, an incorrect geometry results in a bias in the reconstruction of the event vertex of up to 20~mm in the $z$ direction and 40~mm in the $\rho$ direction. By implementing the predicted geometry using our developed model, these biases can be mitigated. With the JUNO experiment now in its data-taking phase, the detector is instrumented with 17,596 20-inch and 25,587 3-inch PMTs~\cite{JUNO:2025fpc}. The current study is primarily limited to the spatial displacements of the PMTs. In practice, the overall detector geometry may also be affected by several other factors, such as the buoyancy-induced deformation of the acrylic sphere. Given the significant challenges in directly measuring these secondary effects, future work might explore the feasibility of using calibration data for data-to-MC comparisons, which could potentially help to quantify these structural uncertainties.

\bibliographystyle{unsrt}
\bibliography{sample}
\end{document}